\documentclass{JHEP3}

\usepackage{lmodern}
\usepackage[T1]{fontenc}

\usepackage{amsmath,amssymb}
\usepackage[dvips]{graphics}
\def\beq{\begin{equation}}
\def\eeq{\end{equation}}
\def\bea{\begin{eqnarray}}
\def\eea{\end{eqnarray}}

\title{Polyakov formulas for GJMS operators from AdS/CFT}
\author{Danilo E. D\'{\i}az
\\ Humboldt-Universit\"at zu Berlin, Institut f\"ur Physik
\\Newtonstr.15, D-12489 Berlin
\\E-mail: \email{ddiaz@physik.hu-berlin.de}}

\abstract{ We argue that the AdS/CFT calculational prescription
for double-trace deformations leads to a holographic derivation
of the conformal anomaly, and its conformal primitive, associated
to the whole family of conformally covariant powers of the
Laplacian (GJMS operators) at the conformal boundary. The bulk side involves a quantum 1-loop
correction to the SUGRA action and the boundary counterpart accounts
for a sub-leading term in the large-N limit. The sequence of GJMS
conformal Laplacians shows up in the two-point function of the CFT operator
dual to a bulk scalar field at certain values of its scaling
dimension. The restriction to conformally flat boundary metrics reduces
the bulk computation to that of volume renormalization which renders
the universal type A anomaly. In this way, we directly connect two chief roles of
the Q-curvature: the main term in Polyakov formulas on one hand, and
its relation to the Poincare metrics of the
Fefferman-Graham construction, on the other hand. We find agreement with
previously conjectured patterns including a generic and simple formula for the
type A anomaly coefficient that matches all reported values in the
literature concerning GJMS operators, to our knowledge.}

\preprint{HU-EP-07/66}
\begin{document}
\section{Introduction}

Conformally covariant differential operators have been the subject
of a continuous interplay between physics and mathematics ever since
the discovery of conformal invariance of Maxwell's equations by
Cunningham~\cite{Cun10} and Bateman~\cite{Bat10} in the early part
of last century. To this early list belongs the Dirac operator,
after Pauli's proof of the conformal invariance of the massless
Dirac equation~\cite{Pau40}, as well as the conformal wave operator,
both in curved spacetime. The Riemannian variant of the later, the
conformal Laplacian, is best known to mathematicians for its role
in the Yamabe problem of prescribing scalar curvature on a
Riemannian manifold~\cite{Yam60}.

In the early 80's a fourth-order conformal covariant was found by
Paneitz~\cite{Pan83}, and independently by Eastwood and
Singer~\cite{ES85}, in relation with gauge fixing Maxwell equations
respecting conformal symmetry. It was rediscovered by
Riegert~\cite{Rie84} while pursuing a different goal, namely a
four-dimensional analog of Polyakov formula~\cite{Pol81} for the
conformal (or trace, or Weyl) anomaly, i.e. a non-local
covariant action whose conformal
variation leads to the general form of the anomaly. Graham, Jenne,
Mason and Sparling~\cite{GJMS92} further showed that the conformal
Laplacian and the Paneitz operator generalize to a family of
conformally covariant differential operators $P_{2k}$ of even order
$2k$, with leading term $\Delta^k$, whenever the dimension $d$ of the
manifold is odd or $d\geq2k$. These `conformally covariant  powers
of the Laplacian' (``GJMS'' operators in what follows) $P_{2k}$ were
obtained using the Fefferman-Graham ambient metric~\cite{FG85}, a
chief tool for the systematic construction of conformal
invariants\footnote{In a physical setting, see the recent
work~\cite{MT06,MMM07} for an alternative route.}.

The feature of the GJMS operators that we will treat in this paper
is the conformal variation of their functional
determinant\footnote{In the mathematical literature, the
zeta-regularized functional determinant is usually meant.} encoded
in a {\em (generalized) Polyakov formula}. In other words, we focus
on the conformal anomaly, and its {\em conformal
primitive}, associated to this family of operators. These formulas
can be worked out case by case in low dimensions from heat kernel
coefficients whose complexity grows significantly with the dimension of
the compact manifold $\mathcal{M}$. However, Branson~\cite{Bra93} succeeded in
finding a pattern in terms of the {\em Q-curvature}, in the
conformally flat case, to rewrite `more invariantly' the quotient of
functional determinants of a conformally covariant operator $A$ (or a
power thereof and with suitable positive ellipticity properties) at
conformally related metrics
$\widehat{g}=e^{\,2w}g$ in any even dimension $d$:
\begin{subequations}
\begin{align}
-\mbox{log}\frac{\mbox{det}\,\widehat{A}}{\mbox{det}
\,A}&=c\int_{\mathcal{M}}
w(\widehat{Q}_d\, dv_{\hat{g}}+Q_d\, dv_g)+\int_{\mathcal{M}}
(\widehat{F}\, dv_{\hat{g}}- F\,dv_g)+(\mbox{global\;term})\\\nonumber\\
&=2\,c\int_{\mathcal{M}}
w(\,Q_d + \,\frac{1}{2}\,P_d\,w)\,dv_g+\int_{\mathcal{M}}
(\widehat{F}\, dv_{\hat{g}}- F\,dv_g)+(\mbox{global\;term})
\quad.\end{align}\label{Poly}\end{subequations}
Here $F$ stands for a local
curvature invariant and the global
term is related to the null space (kernel) of the operator; both vary
depending on what $A$ is, whereas the universal part is captured by the
Q-curvature term. Away from conformal flatness, the pattern is
preserved in $d=2,4,6$ and conjectured to hold in all even
dimension. A related conjecture by Deser and Schwimmer~\cite{DS93}
expresses the infinitesimal variation (anomaly) as a combination of
the Euler density (or {\em Pfaffian}), a local conformal invariant
and a total derivative\footnote{See~\cite{AleI} and ~\cite{BouI}
for independent proofs. The restriction to the conformally flat
class, where the Weyl tensor vanishes, was anticipated
in~\cite{BGP95}.}; but it can be rephrased in terms of the
Q-curvature instead of the Pfaffian.\\
It was in this context that Branson first defined the Q-curvature, in
general even dimension, from the zeroth order (in derivatives) term of the
GJMS operators via analytic continuation in the dimension. The
{\em linear} transformation law under conformal rescaling $\widehat{g}=e^{\,2w}g$
of the Q-curvature generalizes that of the scalar curvature in two
dimensions, \beq e^{d\,w}\,\widehat{Q}_d=Q_d+P_d\,w\;, \eeq and
integrates to a multiple of the Euler characteristic on conformally
flat manifolds.

Another context in which the Q-curvature plays a central role arises in the
volume renormalization of conformally compact asymptotically Einstein
manifolds~\cite{Gra99}, i.e. manifolds whose filling metric is the
Poincare metric of the Fefferman-Graham construction. The renewed
interest in the seminal work of Fefferman and Graham~\cite{FG85}, as an
outgrowth of the relation between the geometry of hyperbolic space (Lobachevsky space)
and conformal geometry on the sphere at the conformal infinity, has been
triggered by the AdS/CFT Correspondence in
physics~\cite{Malda,GKP98,Wit98}. The reconstruction of a {\em bulk
metric} associated to a given {\em conformal structure} at the
conformal infinity~\cite{GL91} and the subsequent evaluation of the
Einstein action with a negative cosmological term becomes the geometrical
task in the limit in which the effective description of string
theory is featured by the classical supergravity approximation. When
the bulk metric is Einstein, the Lagrangian factorizes and the
challenge consists in the regularization of the infinite ({\em
infrared divergent}) volume. A key feature of the AdS/CFT duality is
the fact that infrared divergences in the bulk are related to
ultraviolet ones on the boundary theory, the so called {\em IR-UV
connection}~\cite{SW98}. A further elaboration thereof leads to
the mapping of the conformal anomaly of the gauge theory on the
boundary, at large $N$ (rank of the gauge group) and large 't Hooft
coupling, to the failure of the renormalized bulk action to be
independent of the conformal representative of the metric at the
boundary. This analysis was thoroughly carried out by Henningson and
Skenderis~\cite{HS98}. The {\em holographic anomaly} associated to
the volume is given by the coefficient $v^{(d)}$ of the volume expansion
and its integral gives the coefficient $\mathcal{L}$ of the log-divergent term
in the volume asymptotics. The Q-curvature enters here after the
observation by Graham and Zworski~\cite{GZ03} that the ``integrated
anomaly'' $\mathcal{L}$ is also proportional to the integral of the
Q-curvature. A further refinement by Graham and Juhl~\cite{GJ07}
results in a {\em holographic formula} for the Q-curvature in terms
of the coefficients of the volume expansion.

These later developments involve scattering theory for Poincare
metrics associated to the conformal structure as an alternative
route to GJMS operators and to Q-curvature~\cite{GZ03,FG02}. They are
influenced by, and at the same time generalize, results originally discussed
in the physical context by Witten~\cite{Wit98} for the ``rigid'' case of hyperbolic space.
They are also closely related to the AdS/CFT computation of matter conformal anomalies, where
the powers of the Laplacian that arise in the rigid case~\cite{PS99,dHSS00,PS04} naturally generalize to
the GJMS operators, as outlined in~\cite{dHSS00,PS04}.
In particular, the GJMS Laplacians show up as residues of the scattering operator $S_M(\lambda)$ at
the poles $\lambda=d/2+k$, $k\in\mathbb{N}$. The ``rigid'' version of the
scattering operator $S_M(\lambda)$ is the two-point function $\langle
O_{\lambda}\,O_{\lambda}\rangle$ of a CFT operator $O_{\lambda}\,$ of conformal
dimension $\lambda$ on $\mathbb{R}^d$ or $\mathbb{S}^d$ as conformal
boundary of the half-space model or of the ball model of the hyperbolic
space $\mathbb{H}^{d+1}$, respectively.

In the mapping of anomalies at leading large N, the coupling constant
regimes in which the bulk and boundary computations are done do not overlap;
an underlying non-renormalization of the coefficient of the Euler term (and, therefore, of the Q-curvature)
in the anomaly at leading large $N$ supports the successful matching in two dimensions.
The same is true for the coefficients of the Euler and Weyl terms in four dimensions, where the free field
computation on the boundary involves a combination
of functional determinants of Laplacians on forms, depending on the field content of the multiplet.
However, in six dimensions the coefficient of the Euler density is no longer
protected and agreement between the holographic and the free CFT anomaly computation
is found only for the Weyl terms~\cite{BFT00}. Moreover, the $AdS_9$ holographic conformal anomaly~\cite{NOO00}
is still waiting for a $CFT_8$ description. In consequence, there seems to be little hope to get, via a holographic
procedure involving the classical SUGRA action, the anomaly associated to individual differential
operators (e.g. conformal Laplacian) in generic even dimension $d$, let alone the full determinant, and to
directly connect the Q-curvature terms that naturally arise in both contexts.

In this note, we propose a heuristic ``holographic'' derivation of
the Polyakov formulas for the GJMS operators in the conformally flat
class~(\ref{Poly}). It is based on a remarkable prediction of AdS/CFT
Correspondence, verified in the ``rigid'' case of hyperbolic space as
bulk metric~\cite{GM03,GK03,Gub04,HR06,DD07}, relating corrections to the
partition functions due to a relevant {\em double-trace deformation} of the CFT,
namely a quantum 1-loop in the bulk  and a next-to-leading contribution in
the large-N expansion at the
boundary.
The general situation will involve $(X; g_+)$ as a $d+1$ dimensional
manifold with a Poincare metric and $(M; [g])$ as its
conformal infinity. Our working formula will then be the
natural generalization of the formal equality that was shown to be
valid in dimensional regularization in the rigid
situation~\cite{DD07}: \beq
{\mbox{log}\;\frac{\mbox{det}_{+}[\Delta_X-\lambda(d-\lambda)]}{\mbox{det}_{-}
[\Delta_X-\lambda(d-\lambda)]}} \, =\, -\,{\mbox{log}\,\mbox{det}
\;S_M(\lambda)}\;.\eeq
The determinants of the positive Laplacian on the bulk $X$
are evaluated using the Green's function method, involving the
resolvent at $\lambda$ for the `$+$branch' and its analytic continuation at
$d-\lambda$ for the `$-$branch'. The continuation in the spectral parameter
$\lambda$ to $\lambda=d/2+k$, $k\in\mathbb{N}$ as argument of the scattering
operator $S_M$ on the compact boundary $M$ renders then the functional
determinant of the GJMS conformal Laplacians. The crucial point that
simplifies our present computation is that when restricted to the
conformally flat class, i.e. metrics on $M$ conformal to the
standard round metric on $\mathbb{S}^d$, the bulk $X$ remains (isometric to)
the hyperbolic space $\mathbb{H}^{d+1}$~\cite{GL91,And04}. In this case,
the volume of the hyperbolic space factorizes and the task is then
reduced to volume renormalization, with the only restriction of conformal
flatness of the boundary metric. We focus on the conformal anomaly,
which is then traced back to the holographic anomaly of the
renormalized volume. The usual Hadamard regularization of the volume
produces an anomaly and a corresponding Polyakov formula
which differs from those obtained by standard zeta-regularization on the
boundary, but the {\em universal} type A anomalous term associated to the Pfaffian or to
the Q-curvature does agree.

The paper is organized as follows: we start with the physical
motivation for the functional determinant identity coming from the
generalized AdS/CFT prescription to treat double-trace deformations of the
boundary conformal theory. We then review the rigid case determinants in the
light of several possible regularization techniques. Next we state the conjectured
equality between functional determinants in the general case of a filling Poincare
metric with prescribed conformal infinity. Explicit computations are then presented
in the case of conformally flat boundary metrics for both infinitesimal and finite
conformal variations.  For comparison, we collect then all reported Polyakov formulas
for GJMS operators in the literature, to our knowledge. We finally
conclude by summarizing our holographic findings and hint at possible
further extensions. Some background material is collected in three appendices.
\section{The physical motivation: AdS/CFT correspondence}

The celebrated Maldacena's conjecture~\cite{Malda} and its calculational
prescription~\cite{GKP98,Wit98} entail the equality between the partition
function of String/M-theory (with prescribed boundary conditions) in
the product space $AdS_{d+1}${\tiny$\times$}$Y$, for some compact space $Y$,
and the generating functional of the dual
$CFT_d$ at the conformal boundary. One of the most remarkable successes of this
duality is the mapping of the conformal anomaly at leading large
$N$~\cite{HS98}, as an outgrowth of the IR-UV connection~\cite{SW98}, that
relates the classical supergravity (SUGRA) action in the bulk to a quantum
one-loop anomaly on the boundary.

The rank $N$ of the gauge group measures the size of the geometry in Planck
units ($L_{AdS}/l_P=N^{1/4}$), implying that quantum corrections to this
classical SUGRA action correspond to subleading
terms in the large $N$ limit of the CFT. At this level, there is a universal
AdS/CFT result, not relying on SUSY or any other detail encoded in the compact space
$Y$, concerning an $O(1)$ correction to the conformal anomaly under
a flow produced by a double-trace deformation. This correction was first
computed in the bulk of AdS~\cite{GM03} and confirmed shortly after
by a field theoretic computation on the dual boundary
theory~\cite{GK03} (see also~\cite{Gub04,HR06}). Full agreement was
finally shown with the help of dimensional regularization in
\cite{DD07}, where we were able to match the anomaly as well as the renormalized
values of the functional determinants involved. In the rest of this section
we briefly survey these preliminary results.

\subsection{The generalized prescription for double-trace deformations}

A subtle example of the duality involves a scalar field $\phi$ with ``tachyonic'' mass in the window $-\frac{d^2}{4}\leq
m^2<-\frac{d^2}{4}+1$, first considered long ago by Breitenlohner and Freedman~\cite{BF82}.
Two AdS-invariant quantizations are known to exist, since one may fix either the faster or slower falloff of
the quantum fluctuations of the scalar field at infinity.
The modern AdS/CFT interpretation~\cite{KW99} assigns the same bulk
theory to two different dual CFTs in which the field $\phi$ is dual to an
operators of dimension $\lambda_-$ and $\lambda_+$, respectively, and whose generating
functionals are related to each other by Legendre transformation at
leading large N. The conformal dimensions of the dual CFT
operators, given by the two roots $\lambda
_{\pm}=\frac{d}{2}\pm\nu$ (with $\nu=\sqrt{\frac{d^2}{4}+m^2}$) of the
AdS/CFT relation $m^2=\lambda(\lambda-d)$, are then both above the unitarity bound.

The generalized AdS/CFT prescription to
incorporate boundary multi-trace operators~\cite{Multi} provides a dynamical picture:
a boundary condition on the bulk scalar relating linearly the faster falloff part to the slower one
corresponds to a double-trace deformation of the CFT Lagrangian.
The two CFTs of above are then the end points of a RG flow triggered by the relevant
perturbation $f\,O_{\alpha}^2$ of the $\alpha-$CFT, where the
operator $O_{\alpha}$ has dimension $\lambda_-$ .
The $\alpha-$theory in the UV flows into the $\beta-$theory in the IR, which now has an
operator $O_{\beta}$ with dimension $\lambda_+=d-\lambda_-$ conjugate
to $\lambda_-$. The rest of the operators remains untouched at
leading large $N$, which from the bulk perspective suggests that the metric and the rest of
the fields involved should retain their background values, only the
dual bulk scalar changes its asymptotics\footnote{The simplest
realization of this behavior being the $O(N)$ vector model in
$2<d<4$, see e.g.~\cite{KP02,DeD06}.}.

\subsection{Bulk one-loop effective actions}

Since the only change in the bulk is in the asymptotics of the scalar field, and
pure AdS with zero scalar field remains a solution of the theory at the
classical level for arbitrary linear boundary conditions on the scalar field,
the effect on the partition function cannot be seen at the classical
gravity level in the bulk, i.e. at leading large $N$. However,
the contribution of the quantum fluctuations of the scalar field, given by the functional
determinant of the kinetic term (inverse propagator), are certainly
sensitive to the boundary conditions (reminiscent of the Casimir effect).
In particular, at the endpoints of this RG flow there are two different
propagators $G_{\lambda_-}$ and $G_{\lambda_+}$ corresponding to the two different
AdS-invariant quantizations by fixing the faster or the slower falloff, respectively. The
partition function including the one-loop back-reaction of the scalar field is given by \beq
\label{det-bulk}
Z^{\pm}_{grav}=Z^{class}_{grav}\;\cdot \left[ \mbox{det}_{\pm}
(-\Box + m^2) \right]^{-\frac{1}{2}},\eeq where $Z^{class}_{grav}$
refers to the saddle point approximation, i.e. the classical action.
Using the Green's functions to compute the functional determinants, one realizes that no UV infinities
show up in the ratio $Z^{+}_{grav}/ Z^{-}_{grav}$, since the UV-divergences can be controlled exactly in the same way for both
propagators. In addition, due to the homogeneity of $AdS$, the volume is factorized so that the only divergence
in the ratio of one-loop corrected partition functions is the IR one given by the infinite volume of $AdS$.

The situation is now analog to the leading large N computation of the CFT conformal anomaly,
where the classical Lagrangian density factorizes and is responsible for the $N^2$ factor,
and the geometric part is produced by the regularization of the IR-divergent volume of the bulk.
In consequence, from the above correction to the classical gravitational action (relative change
of the effective cosmological term) one can read off  an  $O(1)$ contribution to the
integrated {\em holographic conformal anomaly}, or equivalently to the CFT central charge, as predicted
by Gubser and Mitra~\cite{GM03}.

\subsection{Boundary partition function}

The analysis of the corresponding effect on the boundary, as done by Gubser and
Klebanov~\cite{GK03},  starts by turning on the deformation
$f\,O_{\alpha}^2$ in the $\alpha-$theory.
Then the Hubbard-Stratonovich transformation (i.e. auxiliary field
trick) can be used to linearize in $O_{\alpha}$ \beq\langle
e^{-\frac{f}{2}\int O_{\alpha}^2}
\rangle\sim\int\mathcal{D}\sigma\,e^{\frac{1}{2f}\int\sigma^2}\,\langle
e^{\int\sigma O_{\alpha}}\rangle~.\eeq Now the large-N
factorization, which means that the correlators are dominated by the
product of two-point functions, is explicitly used to write
\beq\langle e^{\int\sigma O_{\alpha}}\rangle_{N\gg1}\approx
e^{\frac{1}{2}\int\int\sigma\langle
O_{\alpha}O_{\alpha}\rangle\sigma}~.\eeq Finally, integrating back
the auxiliary field produces its fluctuation determinant
$\Xi^{-1/2}=(\frac{\mathbb{I}}{f}+\langle
O_{\alpha}O_{\alpha}\rangle)^{-1/2}$. The
$\beta-$CFT is reached in the limit $f\rightarrow \infty$, so that modulo
unimportant constant factors
\beq
\label{det-bound}
Z_{\beta}/Z_{\alpha}= \left[ \mbox{det}\,\langle
O_{\alpha}O_{\alpha}\rangle\right]^{-\frac{1}{2}}=\left[
\mbox{det}\,\langle O_{\beta}O_{\beta}\rangle\right]^{\frac{1}{2}}~.\eeq

Gubser and Klebanov~\cite{GK03} were able to isolate the coefficient of the
Euler term in the conformal anomaly of the above functional determinant on the round d-sphere.
Explicit computations for several values of the dimension ($d=2,4,6,8$) produced
the same polynomials in $\nu$ as those ``holographically'' predicted by the $AdS_{d+1}$ computation.

\subsection{The functional determinants}

The AdS/CFT correspondence claims the equality between the partition functions. In particular, at
the one-loop level in the bulk and at the corresponding subleading large-N order on the boundary, this
implies the equality between the functional determinants (\ref{det-bulk}) and (\ref{det-bound}) involved in
$Z^{+}_{grav}/ Z^{-}_{grav}=Z_{\beta}/Z_{\alpha}$. As supporting evidence, Hartman and
Rastelli~\cite{HR06} gave a ``kinematical'' explanation. But, as
usual in AdS/CFT correspondence, the prediction is a formal relation between
divergent quantities which has to be properly regularized to make sense out of it.
The only success so far was the correct mapping of the integrated trace
anomaly\footnote{We have been a little cavalier here since the
Breitenlohner-Freedman analysis is done in Lorentzian signature.
However, for computational purposes it is easier to consider the
Euclidean formulation of the CFT and the volume renormalization with
Riemannian signature, so that a Wick rotation should be performed.
The Feynman propagator for the regular modes ($\lambda_+$) in
$AdS_{d+1}$ becomes the resolvent in $\mathbb{H}^{d+1}$, whereas the
continuation to hyperbolic space of the propagator for the irregular
modes is only achieved via the continuation of the resolvent from
$\lambda_+$ to $\lambda_-$~\cite{Cam91}.}.

Finally, in~\cite{DD07} Dorn and the present author were able to find full
agreement between the dimensionally regularized functional determinants for
generic even and odd dimension $d$. Dimensional regularization probed to be a
sensible scheme that puts bulk and boundary divergences on equal footing.
We extended the mapping from that of the integrated anomaly to the renormalized
partition functions as well (cf. appendix~\ref{DR}). The anomaly can be read as the residue of the
pole term. It is important to emphasize that to correctly reproduce the boundary answer,
the contribution from the renormalized volume $\mathcal{V}_{d+1}$
is not enough and there is an additional term, non-polynomial in $\nu$,
multiplying the integrated anomaly $\mathcal{L}_{d+1}$ of the renormalized volume.
This means that the ``naive'' volume factorization is not quite correct, as
explained in~\cite{DD07}, and some combination of UV-vanishing terms from the effective
potential and IR-divergent ones from the volume is needed.
However, due to the conformal invariance of $\mathcal{L}_{d+1}$, we can ignore this
additional term as long as we are interested in the variation under conformal
transformations as in the case of the Polyakov formulas.

\section{Digression on regularizations}

The successful mapping of the bulk and boundary computations in dimensional
regularization is guaranteed by the common regularization tool. It remains
unclear which regularization on the boundary corresponds to the usual cutoff (Hadamard)
regularization of the volume in the bulk. If the regularization schemes of the bulk and
boundary determinants are not on equal footing, only the universal terms in the
anomaly (type A and B) are guaranteed to coincide.
Before embarking ourselves into the study of the behavior under conformal transformations,
let us first illustrate the rigid computation with different regularization
schemes in the simple case of the (positive) Laplacian\footnote{An immediate check of the
connection with GJMS operators comes from the direct evaluation
of the eigenvalue $\frac{\Gamma(l+\frac{d}{2}+\nu)}{\Gamma(l+\frac{d}{2}-\nu)}$ of the intertwiner
(see \ref{eigen}) at integer values
$\nu=k$. These are precisely the eigenvalues of the GJMS operator
$P_{2k}$ on the standard sphere (cf. theo.2.8(f) in~\cite{Bra93}, or~\cite{Gov05}).}
in two dimensions ($d=2$, $\nu=1$).
The raw determinant, involving the eigenvalues
$l(l+1)$ with multiplicity $2l+1$ is simply given by
\beq
\mbox{log det} \Delta = \mbox{tr log} \Delta =\sum_{l=1}^{\infty} (2l+1)\;\mbox{log}
\,[l(l+1)]~,\eeq
where we have excluded the zero eigenvalue corresponding to constant functions.\\

\begin{center}
\bf{\em Bulk and boundary: dimensional regularization}
\end{center}

The full agreement between bulk and boundary computations using dimensional regularization
$D=2-\epsilon$ as shown in~\cite{DD07}, results in a divergent term and a finite remnant
as $\epsilon\rightarrow 0$, given by (\ref{DR1}) and (\ref{DR2}) respectively.
As $\nu\rightarrow 1$, there is an additional divergence due to the zero mode but it happens
on both sides of the equality and can be removed by starting the sum at $l=1$.
After this cancelation and within a ``minimal subtraction'' prescription,
where we just drop the pole, we get a renormalized value for the functional determinant

\beq
\mbox{log det} \Delta = -4\zeta'(-1)-\frac{2}{3}\gamma+\frac{1}{6}
~.\eeq

The integrated anomaly on the two sphere can be directly read from the pole term.
Here $\mathcal{L}_3$ happens
to be conformal invariant only at the integer dimension $d=2$. Away from it,
at $D=2-\epsilon$ and under constant rescaling $g\rightarrow\widehat{g}=e^{2\alpha}g$,
it picks up a factor $e^{(D-2)\alpha}=e^{-\epsilon\,\alpha}$ so that in the limit
$\epsilon\rightarrow 0$ the variation has a zero $e^{-\epsilon\,\alpha}-1$ that
cancels exactly the pole and we end up with minus the residue. But the variation
of the finite remnant, the renormalized value, is just minus the variation of the
divergent part that is thrown away in the renormalization prescription.
In all, the contribution of the local curvature invariants to the integrated conformal
anomaly in the present case is just
\beq
\mathcal{L}_3\,\cdot\left[-\int_0^{1} dx\,4x
\,\mathcal{A}_2(x)\right]
=-\frac{2}{3}
~.\eeq
Alternatively, one can use the fact that in the renormalized result (finite remnant)
at $d=2$ only the term proportional to the renormalized volume $\mathcal{V}_3$ is anomalous.
The integrated anomaly corresponding to the renormalized volume is known to be just $\mathcal{L}_3$.
Keeping track on the coefficient, we find again the same result as above.

However, there is still a missing contribution (a global term) from the exclusion of the zero mode,
as $g\rightarrow\widehat{g}=e^{2\alpha}g$ the eigenvalues are also rescaled
$\lambda_l\rightarrow\widehat{\lambda}_l=e^{-2\alpha}\lambda$ so that we must consider:
 \beq
 -2\alpha\,\sum_{l=1}^{\infty}\mbox{deg}(D,l)
 ~.\eeq
 In dimensional regularization, the sum over degeneracies starting with $l=0$ vanishes,
 so that the above renders an additional contribution
 $2\alpha\,\mbox{deg}(D,0)=2\alpha$. In all,
 \beq
\mbox{log} \frac{\mbox{det}\widehat{\Delta}}{\mbox{det}\Delta} =-\frac{2}{3}\alpha +
2\alpha=\frac{4}{3}\alpha~.\eeq
This totally agrees with the Polyakov formula evaluated for a constant rescaling $w=\alpha$.
The first contribution $-\frac{2}{3}\alpha$ comes from the Q-curvature term $2\,c\int_{\mathcal{M}}
w\,Q_n\,dv_g$, which is precisely the {\it conformal index} of Branson and Ørsted~\cite{BØ86},
and the remaining $2\alpha$ corresponds to the global term associated to the zero mode.
From the rigid computation on the sphere we can therefore read the coefficient of
the Q-curvature term (type A anomaly) as well as the global term. We get no information,
however, on the local invariant term $F$ in the Polyakov formula because it is scale invariant and, therefore,
under rigid rescaling doesn't show up.

The corresponding extension of this computation to higher dimensions and to the whole family of GJMS operators is
straightforward in dimensional regularization.\\

\begin{center}
\bf{\em Boundary: zeta-regularization}
\end{center}

The zeta regularization produces directly a renormalized determinant
\beq
\mbox{log det} \Delta = -\zeta'_{\Delta}(0)~,\eeq
in terms of the zeta function on the two-sphere
\beq
\zeta_{\Delta}(s)=\sum_{\lambda>0}{\lambda^{-s}}=\sum_{l=1}^{\infty}
\frac{2l+1}{[l(l+1)]^s}~.\eeq
The above representation is valid for $Re(s)>1$, the analytic continuation
to $s=0$ can be accomplished by rewriting in terms of better studied zeta functions,
such as Riemann and Hurwitz zeta functions. The result in the present case can be
shown to be~\cite{Wei86}
\beq
\mbox{log det} \Delta = -4\zeta'(-1)+\frac{1}{2}~.\eeq
The integrated anomaly can be read in this case from the variation under
rigid rescaling $\mbox{log\,det}\widehat{A}
-\mbox{log\,det}A=-2\alpha\,\zeta_{A}(0)$. The input we need is the zeta function
$\zeta_{\Delta}(0)=-\frac{2}{3}$ to finally get
\beq
\mbox{log} \frac{\mbox{det}\widehat{\Delta}}{\mbox{det}\Delta}=\frac{4}{3}\,\alpha~.
\eeq

Extensions of this computation to higher dimensions can be attacked
with methods similar to those of~\cite{DK03}. The calculations are rather lengthly and
explicit results, to our knowledge, do not cover the Paneitz operator or any other of the higher GJMS
operators.\\

\begin{center}
\bf{\em Boundary: large-eigenvalue cutoff}
\end{center}

Finally, we want to present yet another regularization tool which is simply a cutoff $l_c$
in the sum over eigenvalues. It is physically appealing due to its role regarding the holographic bound
in AdS/CFT~\cite{SW98}:
the IR-UV connection forces the number of cells in the coarse-grained sphere to be $\epsilon^{-3}$; however,
if one instead truncates at $l_c$ the spherical modes, then the number of modes plays now the same role as
the number of cells did before. The number of modes is given by the counting function (whose asymptotics
is in general given by Weyl's asymptotic formula) that grows as $l_c^3$ in this case. In all,
the identification $l_c\cdot\epsilon\sim1$ between the UV-cutoff $l_c$ and the IR-cutoff $\epsilon$ is
enforced by the requirement of (roughly) one degree of freedom per unit Planck area.

To compute $\mbox{log det} \Delta$ we have to consider the finite sum
\beq
\sum_{l=1}^{l_c} (2l+1)\;\mbox{log}
\,[l(l+1)]=4\sum_{l=1}^{l_c} l\;\mbox{log}\,l
+(2l_c+1)\;\mbox{log}
(l_c+1)~.\eeq
The large $l_c$ asymptotics of the remaining sum is determined by the Glaisher-Kinkelin constant $A$
given by
\beq
1^1 2^2 3^3...n^n=n^{n^2/2+n/2+1/12}\,e^{-n^2/2} (A + o(1))~,\eeq as $n\rightarrow \infty$,
where $\mbox{log}A =\frac{1}{12}-\zeta'(-1)$.
The total contribution for the regularized determinant is then
\beq
-4\zeta'(-1)+\frac{7}{3}+(2l_c^2+4l_c+\frac{4}{3})\,\mbox{log}\,l_c -\frac{1}{4}l_c^2~.
\eeq

We want to read now the integrated anomaly from the log-term, but there seems to be new divergent
terms which have no analog in the volume regularization nor in heat kernel asymptotics. For example,
using a ``proper-time'' cutoff $\delta\rightarrow 0$, the heat kernel produces
\begin{align}
\mbox{log det}\Delta&=-\int_{\delta}\frac{dt}{t}\left\{\frac{1}{4\pi t}
\int_{S^2}\mbox{dvol} \left[1+\frac{R}{6}t \right]- dim\, ker \Delta
\right\} + finite\nonumber\\\nonumber\\&=-\frac{1}{2\delta}+ \frac{2}{3}\,\mbox{log}
\,\frac{1}{\delta}+ finite,
\end{align}
and IR-UV connection relates the cutoffs as $\epsilon\sim\sqrt{\delta}$ in rough agreement with
the volume asymptotics.
Although a physical interpretation of these extra log-divergencies is still obscure to us, their structure is simple enough.
They are in fact proportional to the counting function with the order of the operator as proportionality factor,
$2\cdot\sum_{l=1}^{l_c} (2l+1)=2\cdot(l_c^2+l_c)$. We can subtract them so that the two residual divergences left are the
expected ones in view of the identification $l_c\cdot\sqrt{\delta}\sim1$.
In the present case, the integrated anomaly $\frac{4}{3}$ shows up as coefficient of $\mbox{log}\,l_c$ and the
renormalized determinant, given by \beq
-4\zeta'(-1)+\frac{7}{3}~,\eeq correctly reproduces the `most transcendental' part $-4\zeta'(-1)$
which is common to the two previous regularization alternatives.

The above procedure can be (straightforwardly but tediously) adapted to the higher-dimensional
spheres and to the whole family of GJMS. The corresponding asymptotic estimates are determined
by the Glaisher-Kinkelin-Bendersky constants~\cite{Ben33}(see also \cite{CSA}) in this case.\\

\begin{center}
\bf{\em Bulk: Hadamard regularization of the volume}
\end{center}

The cutoff regularization of the volume~\cite{HS98,Gra99} produces a null renormalized volume when evaluated for
the standard metric of $\mathbb{H}_3$, so that the volume asymptotics is given by
\beq
\mbox{Vol}(\{r>\epsilon\})= \frac{\pi}{2\epsilon^2}-2\pi\,\mbox{log}\frac{1}{\epsilon}+o(1)~.\eeq
When multiplied by the effective potential, the Hadamard-regularized volume will correctly reproduce
the anomaly but not the finite remnant, not even the $-4\zeta'(-1)$ piece. A compensation between divergences in the volume
and vanishing terms in the effective potential, that in dimensional regularization conspired to produce the
correct boundary result, is still missing here.

\section{The general case of Poincare metrics}

After this preamble, let us now turn to the main theme.
To relax the rigidity, we
must first consider a generalization~\cite{Wit98} of the AdS/CFT
Correspondence in terms of a $d+1-$dimensional (asymptotically) Einstein manifold $X$
with negative cosmological constant, that has a compactification
consisting of a manifold with boundary $\overline{X}$ whose boundary
points are $M$ and whose interior points are $X$ with a metric $g_+$ on
$X$ that has a double pole near the boundary so that it defines a
conformal structure $[g]$ on $M$, i.e. the bulk metric is that of a conformally compact
(asymptotically) Einstein manifold.
This defines $(X; g_+)$ as a $d+1$ dimensional
manifold with a Poincare metric and $(M; [g])$ as its
conformal infinity. Such $g_+$ is also asymptotically hyperbolic\footnote{In the physics context,
this generalization is usually described by the somehow less rigorous notion of asymptotically (Euclidean) anti-de Sitter metrics.}.

We are then naturally led to the following guess for the functional
determinants involved in the one-loop bulk correction and the corresponding subleading large-N term
on the boundary:
\beq
{\mbox{log}\;\frac{\mbox{det}_{+}[\Delta_X-\lambda(d-\lambda)]}{\mbox{det}_{-}
[\Delta_X-\lambda(d-\lambda)]}} \, =\, -\,{\mbox{log}\,\mbox{det}
\;S_M(\lambda)}\;.\eeq
Alternatively, the evaluation of the bulk determinant using the Green's function method by taking the derivative with respect
to the spectral parameter leads to the following relation in term of the resolvent $R_X(\lambda)$
and its analytic continuation $R_X(d-\lambda)$
\beq
(d - 2\lambda)\,\mbox{tr}\left[ R_X(\lambda) - R_X(d - \lambda)\right]= \mbox{tr}\left[ S_M^{-1}(\lambda)
\frac{d}{d \lambda}S_M(\lambda) \right].
\eeq
An analog relation has been shown to be valid by Guillarmou (\cite{Gui05}, theo. 1.2) for certain
generalized variants of the trace, but for the case in which $d$ happens to be odd and the functional
determinants are conformal invariants.

Unfortunately, the above functional determinants are too difficult to compute in general. Only in very symmetric
situations they can be explicitly computed. In addition, the determinant of the scattering operator as an elliptic
{\it pseudo-differential operator} is a largely unexplored object.
Yet, Polyakov formulas for the ratio of functional determinants at conformally related metrics capture
valuable information; they only fail to account for conformally invariant terms.
To make some progress, we will then be rather interested in the variation under conformal rescaling of the
boundary metric for even $d$ and consider the continuation in the spectral parameter ($\nu\rightarrow k$, $k=1,2,...,d/2$) to make contact
with the GJMS operators $P_{2k}$, which are better known, and with their corresponding Polyakov formulas.
Conformal flatness of the boundary metric is not assumed in the above guess,
so that under conformal rescaling both type A and type B anomalies will be present.
However, in this paper we will restrict to the conformally flat situation
where the bulk computation will be reduced to that of the volume renormalization
as we will next show.

\section{Conformally flat class and volume renormalization}

To read the (infinitesimal) anomaly, one has to be able to compute the variation under
a Weyl rescaling of the boundary metric. We let the boundary metric to be conformally
related to the standard one on the sphere, so that the bulk geometry is still (isometric to)
the hyperbolic space~\cite{GL91,And04}. In this case, the resolvent is explicitly known in
terms of the hypergeometric function and one easily gets

\beq
\left[R_X(\lambda) - R_X(d - \lambda)\right](x,x)= 2\,\mathcal{A}_d(\nu)~,
\eeq
with $\mathcal{A}_{(d)}(\nu)$, essentially the Plancherel measure on hyperbolic space
at imaginary argument, as in (\ref{eff-pot}). There is no dependence on the position due to the
homogeneity of $\mathbb{H}_{d+1}$, and therefore the volume factorizes when taking the trace.
Integrating back in $\nu$, we get for the ``bare'' determinant

\beq \left[\int_0^k d\nu\,2\nu\,\mathcal{A}_d(\nu)\right]\cdot
\int_{\mathbb{H}^{d+1}} d{\it vol}_{g_+}= -\frac{1}{2}\,
\mbox{log\,det}P_{2k}~.\eeq

A renormalized version (in DR) leaves a finite remnant of the IR-divergent
bulk volume, i.e. the renormalized volume $\mathcal{V}_{d+1}$, and additional conformally invariant
terms with a non-polynomial dependence in $k$ which play no role in the analysis of the conformal variation.
The generalized Polyakov formula relating the functional determinants of the GJMS conformal Laplacians at conformally related
metrics $\widehat{g}=e^{2w}g$ in even dimension $d$, up to the global term, is proportional
to the conformal variation of the renormalized volume

\beq -\frac{1}{2}\, \mbox{log}\, \frac{\mbox{det}\,\widehat{P
}_{2k}}{\mbox{det}\,P_{2k}} = \left[\int_0^k d\nu \,2\nu
\,\mathcal{A}_d(\nu)\right]\cdot
\left(\mathcal{\widehat{V}}_{d+1}-\mathcal{V}_{d+1}\right). \eeq

\subsection{The infinitesimal variation: conformal anomaly}

We have now to deal with the conformal variation of the renormalized volume $V_{d+1}$ as a functional
of the boundary metric representative $g$. Up to total derivative terms, we can rely on the well known results
obtained via a radial cutoff. The infinitesimal variation of the Hadamard-renormalized
volume (see e.g.~\cite{Gra99}) is given by the coefficient $v^{(d)}$ of the volume expansion
\beq
\frac{d}{d\varepsilon}\,\mathcal{V}[e^{2\varepsilon w}g]\mid_{\varepsilon=0}=\int_{\mathcal{M}} w\,v^{(d)}\,dv_g~,
\eeq
i.e. a trace anomaly
\beq \frac{2}{\sqrt{g}}\,g^{\mu\nu}\frac{\delta}{\delta g^{\mu\nu}}\, \mathcal{V} [g]=v^{(d)}~.
\eeq
To get the Q-curvature term we make then use of the holographic formula~\cite{GJ07}
\beq 2 c_{d/2}\,Q = v^{(d)} + ...~,\eeq
where $c_k = (-1)^k [2^{2k}(k)!(k - 1)!]^{-1}$ and the ellipsis stands for derivative terms involving
lower coefficients $v^{(k)}$. The universal Type A anomaly of $-\frac{1}{2}\mbox{log det}\,P_{2k}$,
encoded in the Q-curvature term or in the Euler term, is finally given by

\beq\label{typeA} 2\,c_{d/2}\left[\int_0^k d\nu \,2\nu
\,\mathcal{A}_d(\nu)\right]\cdot Q_d~. \eeq

\subsection{Type A holographic anomaly}

In an independent development, the authors of~\cite{ISTY99} were able to work out the type A
holographic anomaly, coefficient of the Euler term, coming from a generic gravitational
action which admits $AdS$ as solution (see also~\cite{ST08} for an alternative derivation). The input needed is the Lagrangian
density evaluated for the $AdS$ metric, i.e. it suffices to examine the rigid situation
(in Euclidean signature) for hyperbolic space. We can now use this general result combined with the one-loop
computation~\cite{DD07} for the rigid case, to get
\beq \left[\int_0^k d\nu \,2\nu
\,\mathcal{A}_d(\nu)\right]\cdot\frac{E_d}{2^d(d/2)!}~. \eeq
Now, keeping track on the normalizations we can translate $E_d=(-2)^{d/2}(d/2)!\,\mbox{Pff}$ to the Pfaffian,
according to the conventions of~\cite{GJ07} and further use the relation between the Pfaffian and the $v^{(d)}$
coefficient in the conformally flat case $v^{(d)}=\frac{(-2)^{d/2}}{(d/2)!} \mbox{Pff}$. We find then full
agreement with the previous result~(\ref{typeA}) obtained using the volume renormalization.

\subsection{Conformal primitive: Polyakov formula}

To obtain the Polyakov formula for the quotient of the functional determinant at conformally related metrics,
we have to find the conformal primitive of the infinitesimal anomaly. This we can readily do in two ways. We can
apply a result by Branson for the conformal primitive of the Q-curvature term, which gives the universal part
$\int_{\mathcal{M}}w(\,Q_d + \,\frac{1}{2}\,P_d\,w)\,dv_g$ in Polyakov formulas.

Alternatively, via the connection between the renormalized volume and scattering
theory~\cite{GZ03} Chang, Qing and Yang~\cite{CQY} have found an explicit expression for
$\mathcal{\widehat{V}}_{d+1}-\mathcal{V}_{d+1}$ as conformal primitive of $v^{(d)}$. It contains the universal part
of above and additional local curvature invariant terms to correctly reproduce the holographic formula relating $Q_d$
and $v^{(d)}$.

We get then, up to the global term, for the finite conformal variation of the functional determinant

\begin{subequations}
\begin{align}
-\mbox{log}\, \frac{\mbox{det}\,\widehat{P
}_{2k}}{\mbox{det}\,P_{2k}}&=c_{(d,k)}\int_{\mathcal{M}}
w(\widehat{Q}_d\, dv_{\hat{g}}+Q_d\, dv_g)+\int_{\mathcal{M}}(\widehat{F}\, dv_{\hat{g}}- F\,dv_g)
\\\nonumber\\
&=2\,c_{(d,k)}\int_{\mathcal{M}}
w(\,Q_d + \,\frac{1}{2}\,P_d\,w)\,dv_g+\int_{\mathcal{M}}(\widehat{F}\, dv_{\hat{g}}- F\,dv_g)~,
\end{align}\end{subequations}

where
\beq\label{c} c_{(d,k)}= 2\,c_{d/2}\left[\int_0^k d\nu \,2\nu
\,\mathcal{A}_d(\nu)\right]=
\frac{(-1)^{d/2}}{(4\pi)^{d/2}\,(d-1)!\,(d/2)!}\int_0^k d\nu \,
(\nu)_{\frac{d}{2}}\,(-\nu)_{\frac{d}{2}}~\eeq
and the curvature invariants in the $F$-term, which are
regularization-scheme dependent, enter with different coefficient as in the conventional zeta-regularized determinants.
There is clearly a mismatch between zeta-regularization on the boundary and Hadamard regularization in the bulk, as we will
see, and only the universal Q-curvature term is correctly reproduced. Any scheme that renders the Einstein-Hilbert action finite (see e.g.~\cite{PS04,Ole06}) is associated to a renormalized volume. A finite (renormalized) bulk action induces an effective action for the conformal mode on the conformal boundary which gives essentially the Polyakov formula; this has been explicitly shown in \cite{Car05} and \cite{ARZ06} where the Liouville and Riegert actions, respectively, have been obtained. The advantage of the result by Chang, Qing and Yang~\cite{CQY}, besides being simpler and compact, is that the local curvature invariants can be explicitly derived and not only their finite variation under conformal rescaling.
\section{Comparison to ``experiment''}

Here we collect all explicit results we are aware of for the GJMS Laplacians, all of them computed
via standard zeta-regularization.
Most of the known Polyakov formulas are due to Branson and collaborators
(see e.g. \cite{Bra93} and references therein), whose main motivation was related to
sharp inequalities and extremal problems for the functional determinant. The particular values of the
coefficients and the local invariants that enter in the formulas
are extracted from the relevant heat invariants, which are rewritten in the basis
of the Q-curvature, the Weyl invariants plus a total derivative.

\begin{center}
\bf{\em Laplacian, d=2}
\end{center}

This case is the best known in physics, the original\footnote{These formulas ought to be called
Polyakov formulas in ``Liouville's form''. The celebrated non-local covariant form in terms of the Green's
function of $P_d$ is obtained after eliminating the conformal factor $w$ by solving the conformal relation
 $\sqrt{g}\,P_d\,w=\sqrt{\hat{g}}\,\widehat{Q}_d-
\sqrt{g}\,Q_d$.} Polyakov formula~\cite{Pol81} for the
effective action (conformal primitive) of the two-dimensional trace anomaly:

\begin{align}\label{Poly2}
-\mbox{log}\frac{\mbox{det}\,\widehat{\Delta}}{\mbox{det}
\,\Delta}&=\frac{1}{24\pi}\int_{\mathcal{M}}
w(\widehat{R}\, dv_{\hat{g}}+ R\, dv_g)\;-\;\mbox{log}\frac{vol(\hat{g})}{vol(g)}
\nonumber\\\nonumber\\
&=\frac{1}{12\pi}\int_{\mathcal{M}}
w(\,R + \,\frac{1}{2}\,\Delta\,w)\,dv_g\;-\;\mbox{log}\frac{vol(\hat{g})}{vol(g)}
\quad,
\end{align}

In two dimensions the Q-curvature is given by the Schouten scalar
which is half the scalar curvature ($Q_2=J=R/2$), and the conformal Laplacian or
Yamabe operator is simply the Laplacian ($P_2=Y=\Delta$).

\begin{center}
\bf{\em Yamabe, d=4}
\end{center}

The result for the Yamabe operator was first obtained by Branson and {\O}rsted~\cite{BØ91},
although the general structure in four dimensions was anticipated by Riegert~\cite{Rie84}.
The input needed is the heat kernel coefficient $a_4(Y)$ (see e.g.~\cite{Gil79}), then restrict
to the conformally flat class to read the coefficient of the Q-curvature as well as the local
curvature invariants entering in the Polyakov formula to finally write:
\bea\label{Poly4}
-\frac{1}{2}(4\pi)^2\,\mbox{log}\frac{\mbox{det}\,\widehat{Y}}{\mbox{det}
\,Y}&=&-\frac{1}{180}\int_{\mathcal{M}}
w(\widehat{Q}\, dv_{\hat{g}}+ Q\, dv_g)
-\frac{1}{90}\int_{\mathcal{M}}
(\hat{J}^2\, dv_{\hat{g}}- J^2\, dv_g)
\nonumber\\\nonumber\\&=&-\frac{1}{90}\int_{\mathcal{M}}
w(\,Q + \,\frac{1}{2}\,P\,w)\,dv_g
-\frac{1}{90}\int_{\mathcal{M}}
(\hat{J}^2\, dv_{\hat{g}}- J^2\, dv_g)\quad,\eea
where $Q$ is the four-dimensional Q-curvature, $P$ is the Paneitz operator and $J=\frac{R}{2(d-1)}$ is
the Schouten scalar.

\begin{center}
\bf{\em Paneitz, d=4}
\end{center}

For the Paneitz operator in four dimensions, the corresponding Polyakov formula was obtained by
Branson~\cite{Bra96} and the heat invariant input needed was computed from Gilkey's work~\cite{Gil80},

\bea\label{P4}
-\mbox{log}\frac{\mbox{det}\,\widehat{P}}{\mbox{det}
\,P}&=&
\frac{1}{720\pi^2}\left[14\int_{\mathcal{M}}
w(\widehat{Q}\, dv_{\hat{g}}+ Q\, dv_g)
-32\int_{\mathcal{M}}
(\hat{J}^2\, dv_{\hat{g}}- J^2\, dv_g)\right]
-\mbox{log}\frac{vol(\hat{g})}{vol(g)}
\nonumber\\\\\nonumber
&=&
\frac{1}{720\pi^2}\left[28\int_{\mathcal{M}}
w(\,Q + \,\frac{1}{2}\,P\,w)\,dv_g
-32\int_{\mathcal{M}}
(\hat{J}^2\, dv_{\hat{g}}- J^2\, dv_g)\right]
-\mbox{log}\frac{vol(\hat{g})}{vol(g)}
\,.\eea

\begin{center}
\bf{\em Yamabe, d=6}
\end{center}

The six-dimensional case for the Yamabe operator was worked out by Branson~\cite{Bra93}.
The starting point is the heat kernel coefficient $a_6$ computed by Gilkey~\cite{Gil79},
restricted to the conformally flat case.

\begin{align}\label{Y6}
-\frac{3\cdot7!\,(4\pi)^3}{2}\,\mbox{log}\frac{\mbox{det}\,\widehat{Y}}{\mbox{det}
\,Y}=5&\int_{\mathcal{M}}
w(\widehat{Q}_6\, dv_{\hat{g}}+ Q_6\, dv_g)
+13\int_{\mathcal{M}}
(|\hat{\nabla}\hat{J}|^2\, dv_{\hat{g}}- |\nabla J|^2\, dv_g)\nonumber\\\\\nonumber
+34&\int_{\mathcal{M}}
(\hat{J}^3\, dv_{\hat{g}}- J^3\, dv_g)
-32\int_{\mathcal{M}}
(\hat{J}|\hat{V}|^2\, dv_{\hat{g}}- J|V|^2\, dv_g)
\,.\end{align}
Here the local invariant involves now the Schouten tensor $V=\frac{Ric-Jg}{d-2}$ as well.

\begin{center}
\bf{\em Yamabe, d=8}
\end{center}

The eight-dimensional case for the Yamabe operator was worked out by Branson and
Peterson~\cite{BP}, where the necessary input was computed
from Avramidi's result~\cite{Avr86} for the relevant heat invariant. At this stage,
the computation becomes almost prohibiting and it was only done with computer
aided symbolic manipulations\footnote{I am indebted to
L.~J.~Peterson for providing the coefficient of the Q-curvature term and valuable explanations.}
\beq\label{Y8}
-\frac{(4\pi)^4}{2}\,\mbox{log}\frac{\mbox{det}\,\widehat{Y}}{\mbox{det}
\,Y}=-\frac{23}{1360800}\int_{\mathcal{M}}
w(\widehat{Q}_8\, dv_{\hat{g}}+ Q_8\, dv_g)+...
\,.\eeq

\subsection{Further data for the conformal Laplacian}

As far as we are interested in the type A anomaly coefficient, we can make a longer
list based on explicit computations on the spheres via zeta regularization. Once we
know the zeta-function of the operator $\zeta_A(0)$ on the sphere and the dimension
$q(A)$ of its kernel, we can work out the coefficient of the Euler term or,
equivalently, the coefficient of the Q-curvature. The main relation is given by the
{\it conformal index theorem}~\cite{BØ86} restricted to the conformally flat class:
\beq
\zeta_A(0)+q(A)=\frac{c}{l}\int_M  Q_d\, dv_g~,
\eeq
where $2l$ is the order of the differential operator $A$.
There is a vast literature computing the zeta function for the (conformal) Laplacian on
the round sphere. They generalize early results by Weisberger~\cite{Wei86} and are rather
lengthly calculations. Remarkably, there is a compact recipe obtained
by Cappelli and D'Appollonio~\cite{CD'A00} for $d\geq 4$
\beq
\zeta_Y(0)=-\frac{1}{(d-1)!}\int_0^{B(1+B)} dt\,\prod_{i=0}^{d/2-2} [t-i(i+1)],
\eeq
where, after performing the integral, one has to substitute the powers
$B^n$ by the Bernoulli number $B_n$. This produces the results in table~\ref{zeta}:

\begin{table}[ht]
\caption{$\zeta_Y(0)$ on $\mathbb{S}^d$}
\begin{center}
\begin{tabular}{|c|c|c|c|c|}
\hline
& $d=4$ &  $d=6$ &  $d=8$ &  $d=10$\\
\hline
{\large $\zeta_Y(0)$} & $-\frac{1}{90}$ & $\frac{1}{756}$ &
$-\frac{23}{113400}$ & $\frac{263}{7484400}$\\
\hline
\end{tabular}\end{center}
\label{zeta}
\end{table}
In this case, $q(Y)$ is 1 in two dimensions and zero for all other even dimensions
\beq
\zeta_Y(0)+q(Y)=c_Y(d)\, \Gamma(d)\, Vol(S^d)=c_Y(d)\, \Gamma(d)\,
\frac{2\pi^{\frac{d+1}{2}}}{\Gamma(\frac{d+1}{2})}~,
\eeq
so that we can check the coefficient of the Q-curvature for the Polyakov formula for
the conformal Laplacian in any even dimension.

We find agreement with all these values. Our holographic result for the conformal
Laplacian ($k=1$) can be easily compared to the above formula, the integral of
the Q-curvature on $\mathcal{S}^d$ is simply the volume of the sphere times the constant
value $\Gamma(d)$ of the Q-curvature on the round sphere. We get then for the conformal
index an even simpler formula
\beq
\zeta_Y(0)+ q[Y]=\frac{2(-1)^{d/2}}{d!}\int_0^1 d\nu \,(\nu)_{\frac{d}{2}}\,(-\nu)_{\frac{d}{2}}=
\frac{2(-1)^{d/2}}{d!}\int_0^1 d\nu \prod_{i=0}^{d/2-1}[i^2-\nu^2]~.
\eeq

\section{Conclusion}

The main thrust of this paper has been towards a holographic derivation of Polyakov formulas for GJMS operators.
On one hand, this constitutes an important test of the AdS/CFT correspondence in the
general case of a Poincare metric in the bulk. The holographic description of double-trace deformations of the
boundary CFT and the conjectured equality between partition function at the one-loop quantum level in the bulk
and subleading large-N order on the boundary lead to a remarkable identity between functional determinants. We have
been able to make progress in the case of conformally flat boundary metrics, reducing the bulk computation to that
of volume renormalization. This makes, on the other hand, a direct connection between the Q-curvature that appears
in the volume renormalization of Poincare metrics and the universal Q-curvature term in the Polyakov formulas for
conformally covariant operators.
We get a generic formula~(\ref{c}) for the type A conformal anomaly associated to the whole
family of GJMS operators, in agreement with some previously known results that we summarize in table~\ref{exper}.
\begin{table}[ht]
\caption{Q-curvature coefficient $c_{(d,k)}$ in Polyakov formulas for GJMS operators}
\begin{center}
\begin{tabular}{|c|c|c|c|c|c|}
\hline
\multicolumn{6}{|c|}{\large $(d+1)!\cdot (4\pi)^{d/2}\cdot c_{(d,k)}$}\\
\hline
 &  $k=1$, Yamabe &  $k=2$, Paneitz & $k=3$ & $k=4$ & $k=5$\\
\hline
$d=2$ &  $2$ \; \checkmark & - & - & - & -\\
\hline
$d=4$ &  $- 4/3$ \; \checkmark &  $112/3$ \; \checkmark &- & - & -\\
\hline
$d=6$ & $10/3$ \; \checkmark &  $- 64/3$ & $738$ & - & -\\
\hline
$d=8$ &  $- 184/15$ \; \checkmark & $832/15$ & $-1944/15$ & $253184/15$ & - \\
\hline
$d=10$ & $526/9$\; \checkmark  & $-1984/9$ & $1026$ & $-75776/9$ & $4016750/9$ \\
\hline
\end{tabular}\end{center}
\label{exper}
\end{table}

Graham~\cite{Gra99} already noticed that the invariance properties of the renormalized volume
$\mathcal{V}$ are reminiscent of those for the functional determinant of the conformal Laplacian, which is conformally
invariant in odd dimensions but which has an anomaly in even dimensions,
and that the  properties of the invariant $\mathcal{L}$ are, on the other hand, similar to those for the constant term in the
expansion of the integrated heat kernel for the conformal Laplacian, which vanishes in odd dimensions but in even
dimensions is a conformal invariant obtained by integrating a local expression in curvature. In this note, we have gone further
and shown that the above similarities can be promoted to equalities, up to regularization-scheme dependent terms.
The Polyakov formulas obtained via volume renormalization correctly reproduce the universal
Q-curvature term. However, the coefficients of the additional local curvature invariants are different from those obtained
via $\zeta$-regularization on the compact boundary. It thus remains a challenge to find a bulk regularization that
corresponds to the $\zeta$-regularization on the boundary.\\
Notwithstanding, we can unambiguously write down a compact formula for the zeta function of the GJMS
operators on the round sphere
\beq
\zeta_{P_{2k}}(0)+ \delta_{d,2k}=\frac{2}{k}\frac{(-1)^{d/2}}{d!}\int_0^k d\nu
\,(\nu)_{\frac{d}{2}}\,(-\nu)_{\frac{d}{2}}~,
\eeq
which correctly reproduces all values reported in the literature and predicts new ones (table~\ref{zeta-GJMS}).
It would be desirable to have a confirmation of these results; one possible way goes via the relation of the conformal
anomaly with the Wodzicki residue (see e.g.~\cite{PR05}) and the computation of the later using symbol calculus.
\begin{table}[ht]
\caption{$\zeta_{P_{2k}}(0)$ on $\mathbb{S}^d$}
\begin{center}
\begin{tabular}{|c|c|c|c|c|c|}
\hline
 &  $k=1$, Yamabe &  $k=2$, Paneitz & $k=3$ & $k=4$ & $k=5$\\
\hline
$d=2$ &  $-\frac{2}{3}$ \; \checkmark & - & - & - & -\\
\hline
$d=4$ &  $-\frac{1}{90}$ \; \checkmark &  $-\frac{38}{45}$ \;
\checkmark &- & - & -\\
\hline
$d=6$ & $\frac{1}{756}$ \; \checkmark &  $-\frac{4}{945}$ &
$-\frac{379}{420}$ & - & -\\
\hline
$d=8$ &  $-\frac{23}{113400}$ \; \checkmark & $\frac{13}{28350}$
& $-\frac{1}{1400}$ & $\frac{562603}{604800}$ & - \\
\hline
$d=10$ & $\frac{263}{7484400}$ \; \checkmark  & $-\frac{31}{467775}$
& $\frac{19}{92400}$ & $-\frac{592}{467775}$ & $-\frac{283309}{299376}$ \\
\hline
\end{tabular}\end{center}
\label{zeta-GJMS}
\end{table}

We have favored the Q-curvature, rather than the Euler density, to describe the type A anomaly. This is mainly due to
its simpler transformation law under conformal rescaling and its simpler conformal primitive. In conformal geometry, the
Q-curvature certainly plays a central role and has been intensively studied in recent years. On the physical side, it has been less explored;
however, let us mention that some purely QFT considerations regarding the irreversibility of the RG flow,
unitarity and positivity of the induced action for the conformal factor and {\it a-theorem} have led Anselmi~\cite{Ans99,Ans99_2} to
introduce a ``pondered Euler density'' in the study of conformal anomalies. This ``pondered Euler density'' has a linear transformation law under conformal recaling, therefore it is nothing but the Q-curvature modulo Weyl invariant terms. Moreover, the explicit expression for $d=6$ in~\cite{Ans99_2} coincides with Branson's $Q_6$ in six dimensions~\cite{Juh08}.

There are still several interesting issues to be explored. Going beyond conformal flatness
will switch on the Weyl-terms whose number grows with the dimension. Here, the issue of
uniqueness of the filling Poincare metric, together with the topology of the conformal
boundary, will surely play an important role\footnote{In a celebrated example due to Hawking and Page~\cite{HP83}, where
the AdS Schwarzschild black hole and thermal AdS share the
same conformal infinity, the bulk one-loop effect corresponding to the double-trace deformation has already been
explored in~\cite{NO03}.}. Even the rigid case of hyperbolic space can be extended to quotients by symmetry groups, as is
the case of Kleinian groups, where connections with number theory via Selberg zeta functions naturally arise
(see e.g.~\cite{PP01,PW03} and~\cite{Moo04}, sect.2.9, for a related discussion).
In another direction, the relation between Polyakov formulas, extremal of functional determinants
and sharp inequalities~\cite{Bra93} may well admit a holographic interpretation in terms of the bulk geometry.
\acknowledgments

My deep gratitude to Harald Dorn for encouragement and valuable advice in
preparing this manuscript. I benefited from common seminars with the group of
Differential Geometry and Global Analysis at HU-Berlin given by J. Erdmenger
and A. Juhl. I am also indebted to H. Baum, G. Jorjadze, L.J. Peterson and H.S. Yang for useful
discussions and references. Finally, I thank the Floro Pérez Alumni Forum for the encouragement.

\begin{appendix}

\section{GJMS operators and Q-curvature}
\label{Q-cur}

To give a glimpse of these constructions in conformal geometry, let
us go back to the Poincare patch and examine the analytic
continuation to $\lambda>d/2$ of the kernel
\beq\frac{1}{|\overrightarrow{x}'-
\overrightarrow{x}|^{2\lambda}}\,.\eeq
It will have single poles at $\lambda=d/2+k, k\in\mathbb{N}$, since
in the neighborhood of these values (see e.g.~\cite{GS64})
\beq\lim_{\lambda\rightarrow d/2+k}\frac{\lambda-d/2-k}
{|\overrightarrow{x}|^{2\lambda}}=-c_k\;\Delta^k\,\delta^{(d)}
(\overrightarrow{x})\eeq where \beq
c_k=\frac{1}{2^{2k}\,k!\,(k-1)!}.\eeq Therefore for these ``resonant
values'' the action of the kernel reduces to that of the k-th power
of the Laplacian $\Delta^k$, a conformal invariant (covariant)
differential operator\footnote{There is a factor $(-1)^k$ hanging
around, just because in the mathematical literature the {\em
positive Laplacian} is preferred.}.

The generalization of this observation~\cite{GZ03} for a filling
Poincare metric associated to a given conformal structure involves
$P_{2k}$, the conformally invariant operators of GJMS~\cite{GJMS92}.

\begin{center}
{\em GJMS operators}
\end{center}
The GJMS operators $P_{2k}$ built using the Fefferman-Graham ambient
construction have, among others, the following properties in a d-dim
Riemannian manifold $(M,g)$
\begin{itemize}
\item On flat $\mathbb{R}^d$, $P_{2k}=\Delta^k$
\item $P_{2k}\,$ $\exists\,k\in\mathbb{N}$  and $k-d/2\neq
\mathbb{Z}^+$
\item $P_{2k}=\Delta^k+(lower\,order\,terms)$
\item $P_{2k}$ is self-adjoint
\item for $f\in C^{\infty}(M)$, under a conformal change
of metric $\widehat{g}=e^{2\sigma}g$, $\sigma\in C^{\infty}(M)$,
conformal covariance: $\widehat{P}_{2k}
f=e^{-\frac{d+2k}{2}\sigma}P_{2k}(e^{\frac{d-2k}{2}\sigma}f)$
\item $P_{2k}$ has a polynomial expansion in $\nabla$ and the Riemann
tensor (actually the Ricci tensor) in which all coefficients are
rational in the dimension $d$
\item $P_{2k}$ has the form $\nabla\cdot(S_k \nabla) + \frac{d-2k}{2}
\,Q^d_k$,
where $S_k=\Delta^{k-1}+(lower\,order\,terms)$ and $Q^d_k$ is a local scalar invariant.
\end{itemize}

\begin{center}
{\em Q-Curvature}
\end{center}

The $Q-curvature$ generalizes in many ways the 2-dim scalar
curvature $R$. It original derivation tries to mimic the derivation
of the {\em prescribed Gaussian curvature equation} (PGC) in 2-dim
starting from the {\em Yamabe equation} in higher dimension and
analytically continuing to $d=2$.

Start with the conformal transformation of the scalar curvature at
$d\geq3$ \beq e^{2\sigma}\widehat{R}=R+2(d-1)\Delta \sigma -
(d-1)(d-2)\nabla\sigma \cdot\nabla\sigma\eeq
 and absorb the quadratic term
 \beq\Delta \sigma -(d/2-1)\nabla\sigma\cdot\nabla\sigma=\frac{2}{d-2}
 \,e^{-(d/2-1)\sigma}\,\Delta\, e^{(d/2-1)\sigma},\eeq
 to get for the Schouten scalar $J:=\frac{R}{2(d-1)}$ and
 $u:=e^{(d/2-1)\sigma}$ the Yamabe equation
\beq[\Delta+(d/2-1)J]\,u=(d/2-1)\,\widehat{J}\,u^{\frac{d+2}{d-2}}.\eeq
The trick (due to Branson) is now to slip in a $1$ to rewrite as
\beq \Delta(e^{(d/2-1)\sigma}-1)+(d/2-1)\,J\,e^{(d/2-1)\sigma}\,=
\,(d/2-1)\,\widehat{J}\,e^{(d/2+1)\sigma}\eeq
 and take now the limit $d\rightarrow2$ that results in the PGC eqn.
\beq e^{2\sigma}\,\widehat{J}\,=\,J+\Delta\,\sigma.\eeq

The very same trick applied now to the higher-order Yamabe eqn.
based on the GJMS operators \beq P_{2k}\,u\,=\,\nabla\cdot S_k
\nabla\,(u-1)+(d/2-k)\,Q^d_{2k}\,u\,=\,(d/2-k)\,\widehat{Q^d_{2k}}\,
u^{\frac{d+2k}{d-2k}}\eeq with $u=e^{(d/2-k)\sigma}$, in the limit
$d\rightarrow 2k$ renders the higher (even-)dimensional
generalization of the PGC eqn. \beq
e^{d\sigma}\widehat{Q}\,=\,Q+P\,\sigma~,\eeq where $Q:=Q^d_d$
and $P:=P_d$.

Among the properties of the Q-curvature, the conformal invariance of
its volume integral easily follows.

\section{Q-curvature and volume renormalization}
\label{vol}

Let the bulk metric to be that of an conformally
compact (asymptotically) Einstein manifold,
i.e. $Ric(g_+)=-d\,g_+$. The bulk geometry can be partially
reconstructed by an asymptotic expansion, which is essentially the
content of the Fefferman-Graham theorem~\cite{FG85}. One can always
find local coordinates near the boundary (at $r=0$) to write the
bulk metric as \beq g_+=r^{-2}\{dr^2+g_r\}.\eeq Euclidean
$AdS_{d+1}$ corresponds to the choice $g_r=(1-r^2)^2g_0$ with
$4\,g_0$ being the round metric on the sphere $S^d$. The
``reconstruction'' theorem leads to the asymptotics \bea
&d& odd: \nonumber\\
&&g_r=g^{(0)}+g^{(2)}r^2+
(even\,powers)+g^{(d)}r^d+...\\\nonumber\\
&d& even: \nonumber\\
&&g_r=g^{(0)}+g^{(2)}r^2+ (even\,powers)+g^{(d)}r^d+hr^d\mbox{log
}r+...\eea
 where $g^{(0)}=g$ is the chosen metric at the conformal boundary.
For odd $d$, $g^{(j)}$ are tensors on the boundary and $g^{(d)}$ is
trace-free. For $0\leq j\leq d-1$, $g^{(j)}$ are locally formally
determined by the conformal representative but $g^{(d)}$ is formally
undetermined, subject to the trace-free condition. For even $d$,
$g^{(j)}$ are locally determined for $j$ even $0\leq j\leq d-2$, $h$
is locally determined and trace-free. The trace of $g^{(d)}$ is
locally determined, but its trace-free part is formally
undetermined. All this is dictated by Einstein equations.

The volume element has then an asymptotic expansion \bea
dv_{g_+}&=&\sqrt{\frac{det g_r}{det g}}\frac{dv_g\,dr}{r^{d+1}}
\nonumber\\&=&\{1+v^{(2)}r^2+
(even\,powers)+v^{(d)}r^d+...\}\frac{dv_g\,dr}{r^{d+1}},\eea where
all coefficients $v^{(j)}, j=1...d$ are locally determined in term of
curvature invariants of the boundary metric and  $v^{(d)}=0$ if $d$ is
odd.

Before taking the $r-$integral, a regularization is needed. Then a subtraction
(renormalization) prescription renders a finite answer when the regulator is removed.
When $d$ is odd, the finite remnant $\mathcal{V}$ in the expansion
(renormalized volume) turns out to be independent of the conformal
choice of the boundary metric. If $d$ is even, in turn, $\mathcal{V}$ is no
longer invariant and its variation under a Weyl transformaton of the
boundary metric gives rise to the {\em conformal anomaly}. It is
the coefficient $\mathcal{L}$ of the log$\epsilon$-term in the Hadamard (cutoff) regularization or residue at the
pole in dimensional and Riesz regularizations
\beq \mathcal{L}=\int_\mathcal{M} v^{(d)}\,dv_g~,\eeq
given by the integral of a
local curvature expression on the boundary, the
invariant one in this case. The variation of
$\mathcal{V}$ happens to be connected to this invariant:
$g\rightarrow e^{2w}g$ for infinitesimal $w$ makes
$\mathcal{V}\rightarrow \mathcal{V}+\int w\,v^{(d)}\,dv_g$ in the Hadamard
regularization scheme.

The Q-curvature enters here and provides one of the important terms
in volume renormalization asymptotics at conformal
infinity~\cite{GZ03} \beq \mathcal{L}=2\,c_{d/2}\int_{\mathcal{M}}Q_d\,dv_g
~.\eeq Therefore, the Q-curvature is then proportional to the $v^{(d)}$
coefficient in the volume expansion, up to total-derivative terms which are explicitly
given by the ``holographic formula''~\cite{GJ07}.


\section{Rigid case in dimensional regularization}
\label{DR}

The rigid computation in the bulk involves the volume renormalization of the ball model of the
hyperbolic space with the standard metric:
\beq
\label{naive-fact}
-\mbox{log}\,Z^+_{grav}/Z^-_{grav}= \left[\int_0^\nu dx\,2x\,\mathcal{A}_d(x)\right]\cdot
\int_{\mathbb{H}_{d+1}} d{\it vol}~.\eeq
The factor in square brackets comes from the difference of the one-loop effective potentials
associated to the two asymptotic behaviors, whose short distance divergences cancel out to
render a finite result with
\beq\label{eff-pot}  \mathcal{A}_d(\nu)\,
=\,\frac{1}{2\nu}\,\frac{1}{2^d\,\pi^{\frac{d}{2}}}\;
\frac{(\nu)_{\frac{d}{2}}\,(-\nu)_{\frac{d}{2}}}{(\frac{1}{2})_{\frac{d}{2}}}~.\eeq

The boundary computation on the standard sphere $\mathbb{S}^d$, expanding in spherical harmonics,
results in an UV-divergent sum
\beq
\label{eigen}
-2\;\mbox{log}\,Z_{\beta}/Z_{\alpha}=\sum_{l=0}^{\infty}\mbox{deg}(d,l)\;\mbox{log}
\frac{\Gamma(l+\frac{d}{2}+\nu)}{\Gamma(l+\frac{d}{2}-\nu)}~.\eeq
Here we have a weighted sum with the degeneracies of the spherical harmonics
\beq\mbox{deg}(d,l)=\frac{2l+d-1}{d-1}\frac{(d-1)_l}{l!}~,\eeq
and the ratio $\frac{\Gamma(l+\frac{d}{2}+\nu)}{\Gamma(l+\frac{d}{2}-\nu)}$ are nothing but
the eigenvalues of the \textit{intertwiner}~(cf. eq.2.13 in~\cite{Bra93}) between conjugate
representations (with conformal labels $\lambda_-$ and $\lambda_+$), which is the two-point function
$\langle O_{\beta}O_{\beta}\rangle$ on the round sphere (see e.g.~\cite{Dob98}).

We extended the mapping from that of the integrated anomaly to the renormalized
partition functions as well. The anomaly can be read as the residue of the
pole term
\beq\label{DR1} \mathcal{L}_{d+1}\;\cdot \mathcal{A}_d(\nu)~,\eeq
and the renormalized determinant is given by
\beq
\label{DR2}
-\mbox{log}\,Z^+_{grav}/Z^-_{grav}=\left[\int_0^\nu dx\,2x\,\mathcal{A}_d(x)
\right]\;\cdot\mathcal{V}_{d+1}
\;+\;\left[\int_0^\nu dx\,2x\,\mathcal{B}_d(x)\right]\;\cdot\mathcal{L}_{d+1}
~,\eeq
where
\beq
\mathcal{B}_d(\nu)=\frac{\mathcal{A}_d(\nu)}{2}\left\{
\mbox{log}(4\pi)+\psi(\frac{1}{2}-\frac{d}{2})
-\psi(\frac{d}{2}+\nu)-\psi(\frac{d}{2}-\nu)\right\}~.\eeq

\end{appendix}

\end{document}